\documentclass{pasj00}
\draft

\begin{document}
\SetRunningHead{Author(s) in page-head}{Running Head}
\Received{2000/12/31}
\Accepted{2001/01/01}

\title{Title of Your Paper}


\newcommand{\EG}{3EG J1835+5918}
\newcommand{\RX}{RX J1836.2+5925}

\title{A Deep Optical Observation for an Enigmatic 
Unidentified Gamma-Ray Source 3EG J1835+5918
}

\author{Tomonori \textsc{Totani}\altaffilmark{1, 2}, Wataru \textsc{Kawasaki}$^{3}$, and
Nobuyuki \textsc{Kawai}\altaffilmark{4}}

\altaffiltext{1}{Princeton University Observatory, Peyton Hall, \\
Princeton, NJ08544-1001, USA. (e-mail: ttotani@princeton.edu)}
\altaffiltext{2}{Theory Division, 
National Astronomical Observatory, Mitaka, Tokyo 181-8588}
\altaffiltext{3}{Department of Astronomy, The University of Tokyo, \\
7-3-1, Hongo, Bunkyo-ku, Tokyo, 113-0033
}
\altaffiltext{4}{
Department of Physics, Tokyo Institute of Technology, 2-12-1,
            Ookayama, Meguro, Tokyo 152-0033 }


\maketitle

\begin{abstract}
We report a deep optical imaging observation by the Subaru telescope for a
very soft X-ray source {\RX}, which has been suspected to be an isolated
neutron star associated with the brightest as-yet unidentified EGRET source
outside the Galactic plane, {\EG}. An extended source having a complex,
bipolar shape is found at $B\sim 26$, and this might be an extended pulsar
nebular whose flux is about 5-6 orders of magnitude lower than gamma-ray
flux, although finding a galaxy of this magnitude by chance in the error
circle is of order unity.  We have found two even fainter, possibly point
sources at $B \sim 28$, although their detections are not firm because of low
signal-to-noise.  If the extended object of $B \sim 26$ is a galaxy and not
related to 3EG J1835+5918, a lower limit on X-ray/optical flux ratio is set
as $f_X/f_B \gtrsim 2700$, giving a further strong support of the
neutron-star identification of 3EG J1835+5918. Interestingly, if either of
the two sources at $B \sim 28$ is the real counterpart of {\RX} and thermal
emission from the surface of an isolated neutron star, the temperature and
distance to the source become $\sim 4 \times 10^5$K and $\sim 300$pc,
respectively, showing a striking similarity of its spectral energy
distribution to the proto-type radio-quiet gamma-ray pulsar Geminga. No
detection of nonthermal hard X-ray emission is consistent with the ASCA upper
limit, if the nonthermal flux of {\EG}/{\RX} is at a similar level with that
of Gemiga.
\end{abstract}

\section{Introduction}
Over half of GeV gamma-ray sources detected by the EGRET experiment
have not yet been identified as known astronomical objects (Hartman et
al. 1999), and understanding their origin is one of the most important
issues in high-energy astrophysics. Many of unidentified sources
located in the Galactic plane are believed to be associated with
either of pulsars, supernova remnants, or massive stars as suggested
by statistically significant correlation between unidentified EGRET
sources and tracers of these objects (see, e.g., Romero 2001 for a
review). Another Galactic population of unidentified EGRET sources at
intermediate Galactic latitude ($|b| \lesssim 40^\circ$) has been
reported by Gehrels et al. (2000), which are apparently associated
with the Gould belt, and they might be off-beam gamma-ray pulsars
(Harding \& Zhang 2001). There are also unidentified sources at even
higher latitude ($|b| \gtrsim 40^\circ$), suggesting the extragalactic
origin. Variable sources among them are likely undetected blazars,
while some stable sources might be dynamically forming or merging
clusters of galaxies (Totani \& Kitayama 2000; Waxman \& Loeb 2000;
Kawasaki \& Totani 2002).

One of the unidentified EGRET sources, {\EG} has been paid particular
attention in recent years, since it is the brightest unidentified
source outside the Galactic plane ($l = 88.7^\circ$, $b =
25.1^\circ$). Its gamma-ray properties, i.e., steady flux and hard
spectrum are consistent well with other gamma-ray pulsars observed by
EGRET (Reimer et al. 2001), rather than blazars. No strong radio
counterpart also argues against the blazar origin. Intensive
multi-wavelength studies of this source (Mirabal et al. 2000; Mirabal
\& Halpern 2001; Reimer et al. 2001) have revealed about a dozen of
X-ray sources in or around the error circle of {\EG}, and most of them
are coronal emission from stars or quasars without any special or
blazar-like characteristics, which are not likely the counterpart of
{\EG}.  Then the only one unidentified X-ray source, {\RX}, remains as
the most likely counterpart. It has very soft X-ray spectrum ($T < 5
\times 10^5$K) and has no optical counterpart down to $V \sim 25$
(Mirabal \& Halpern 2001) giving a high X-ray/optical ratio. This and
the lack of strong radio emission are reminiscent of the
characteristics of isolated neutron stars or radio-quiet pulsars
(e.g., Brazier \& Johnston 1999; Neuh\"auser \& Tr\"umper 1999), and
this is why both Mirabal \& Halpern (2001) and Reimer et al. (2001)
concluded that the most likely identification of {\EG} / {\RX} is an
isolated radio-quiet pulsar, which is similar to the proto-type object
Geminga.

However, with the X-ray flux of 2--6$ \times 10^{-13} \ \rm erg \
cm^{-2} s^{-1}$ (Mirabal \& Halpern 2001), 
the optical upper limit ($V > 25.2$ at $3\sigma$) corresponds
to the X-ray/optical flux ratio of $f_X / f_{\rm V} \gtrsim$
300, and it is not large enough compared with the flux ratio 
expected for isolated pulsars or neutron stars, although
all X-ray sources other than isolated neutron stars and low-mass
X-ray binaries have $f_X / f_{\rm V} \lesssim 80$ (Stocke et al. 1991).
Here we report a deep optical imaging observation of {\EG} / 
{\RX} by the 8.2m Subaru telescope, to examine the proposed
identification and the origin of this enigmatic source.

\section{The Subaru Observation}
\begin{figure*}
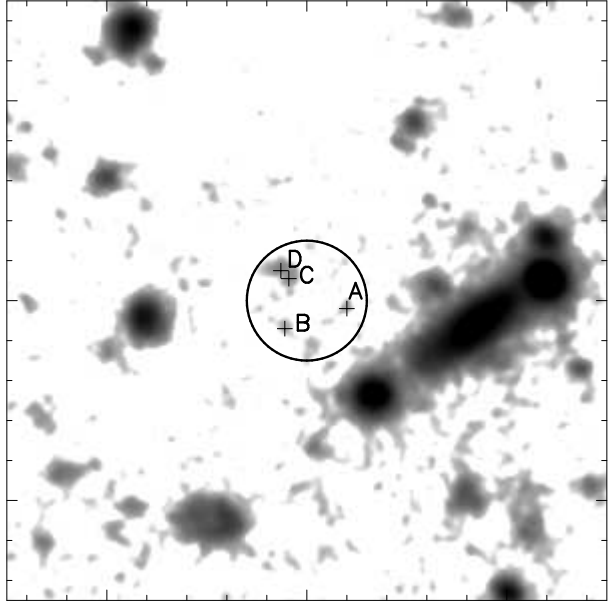

   \begin{center}
      \FigureFile(80mm,50mm){cleanimageB30x30bw.eps}
   \end{center}
\caption{The $30''\times 30''$ image in the $B$ band around the region
of {\RX} whose error circle is shown by the circle. The four objects
found in the circle are indicated as A--D. North is up.}
\label{fig:B-30x30}
\end{figure*}

The observation was made on July 17, 2001, by the Subaru/FOCAS whose field of
view is a circle with $6'$ diameter.  We took images in $B$ and $U$ bands,
with the total exposure time of 2400 and 6000 sec, respectively. The
5$\sigma$ peak level corresponds to $B = 27.5$ and $U= 25.9$. Figures
\ref{fig:B-30x30} and \ref{fig:B-10x10} show the B band images centered on
the position of {\RX} determined by Mirabal \& Halpern (2001) with the error
radius of $3''$ (J2000, $18^h 36^m 13^s.77$, $+59^\circ 25'30''.4$).  The
Galactic extinction to this direction is $A_B = 0.22$ and $A_U = 0.28$
(Schlegel, Finkbeiner, \& Davis 1998) giving upper bounds on extinction in
any Galactic objects. The seeing size in the $B$ band is $0.9''$ FWHM. On the
other hand, the quality of $U$ band image is not as good as the $B$ band and
no object is found in the error circle; we then set a rough upper limit of $U
< 24.8$ including systematic uncertainty of photometry.

In the $B$ band image we detected two extended sources which are connected
with each other (labeled as C and D in the images).  Their peak are 5.1 and
5.8$\sigma$ of the sky fluctuation level, respectively. We found two point-like
sources with peak levels higher than 3$\sigma$ levels, which are labeled as A
and B in the images.  The results of photometry as well as the source
locations for the above four objects are summarized in Table
\ref{table:photometry}. In the following of this paper, we will refer to
the sources C and D as one source (C+D). We checked that the noise
fluctuation is symmetric above and below the zero point of the background,
and we did the same detection procedure for the 
inverted image to estimate the number of spurious objects. This
indicates that sources A and B are
not firm detections; the probabilities that they are spurious detection
are estimated to be $\sim 60$ and 30\%, respectively, from the counts
of spurious objects in the inverted frame.
An even fainter, extended source can be seen in the bottom of the
error circle, whose peak level is less than 3$\sigma$, and we do not consider
this source in this letter.

If {\EG} is a pulsar with high proper motion, it might have moved to outside
of the error circle determined by ROSAT data. Therefore examination only within
the ROSAT error circle is not sufficent.  Our observation is 3.5yr later than
the ROSAT HRI observation (Mirabal \& Halpern 2001), and assuming a distance
of 200pc and transverse velocity of 500km/s, the proper motion becomes $\sim
2''$.  Therefore we examined the outer region $2''$ beyond the ROSAT
error circle, but we found no significant sources like A--D.

\section{Implications and Discussion}

First we consider the possibility that C+D is an interacting galaxy.  The
$B$-band galaxy count at $B \sim 26$ is about $2 \times 10^5 \ \rm mag^{-1}
deg^{-2}$ (e.g., Totani \& Yoshii 2000), and hence the number of galaxies
expected in the error circle of {\RX} becomes $\sim$0.4, indicating that the
appearance of a galaxy by chance is not surprising at all.  The X-ray spectrum
of {\RX} is very soft, and this argues against the possibility that the X-ray
emission is coming from nuclear activity in extended faint galaxies. As
mentioned earlier, the gamma-ray properties and lack of strong radio emission
also disfavor the AGN origin for {\EG}.  The morphology of C and D suggests
that they are merging or interacting galaxies, and steady radiation of
gamma-rays may be possible by particle acceleration from merging shocks.
However, assuming typical values of mass/light ratio of galaxies, merging
velocity ($\sim$ a few hundreds km/s), and interaction time scale ($\sim
10^8$yr), we estimated the maximum luminosity possible from merging,
as the merging kinetic energy divided by the interaction time
scale.  It turns out to be of order $\sim 10^{-14} \rm erg \ cm^{-2} s^{-1}$,
and this is more than four orders of magnitude weaker than the gamma-ray flux
of 3EG J1835+5918, safely excluding this possibility. Therefore, if the source
C+D is a galaxy, it is not a counterpart of {\EG}.

Another possibility which should be considered here is that the extended
object C+D is a pulsar wind nebula. The optical $\nu F_\nu$ flux is more than
5 orders of magnitude lower than the gamma-ray flux, and hence than the
spin-down luminosity of the pulsar.  The size of the extended objects, $\sim
1''$, seems small as an entire size of typical synchrotron nebulae, but
observations of other pulsars or pulsar candidates have often revealed
extended cores or knots with similar size around central pulsars, e.g.,
optical knots in the Crab pulsar (Hester et al. 1995) or Chandra observations
of several Crab-like young supernova remnants (Slane et al. 2000; Hughes et
al. 2001). Nonthermal X-ray synchrotron nebula emission is known to be
correlated to the spin-down luminosity as $L_X \sim 10^{-2} \dot{E}$ (e.g.,
Seward \& Wang 1988; Becker \& Tr\"umper 1997), and the compact (but extended)
X-ray cores have a flux of about 10\% of the total nebula flux (Slane et
al. 2000; Hughes et al. 2001). The optical flux of the object C+D seems small
compared with these X-ray fluxes. On the other hand, the optical knots of the
Crab pulsar have luminosity about seven orders of magnitude lower than the
spin-down luminosity, and hence the object C+D may be a similar knot embedded
in a nebula which is much more extended and undetected.  Since the object is
located at intermediate Galactic latitude ($b=25.2^\circ$), the pulsar is
likely to have a high velocity. Then it is possible that the extended nebular
is a bow shock where the ram pressure from interstellar matter and the pulsar
wind pressure are in equilibrium:
\begin{equation}
n m_p \upsilon^2 = \frac{\dot E}{4 \pi r_s^2 c} =
\frac{f_\gamma^{-1} (4 \pi D^2 F_\gamma)}{4\pi r_s^2 c}
= \frac{f_\gamma^{-1} F_\gamma}{c \theta_s^2} \ ,
\end{equation}
where $\dot E$ is the spin-down luminosity assumed to be the same as
the wind luminosity,
$n$ the nucleon density of ambient medium, $m_p$ the proton mass,
$\upsilon$ the pulsar velocity,
$D$ the distance to the pulsar, $f_\gamma$ the ratio of gamma-ray
luminosity to spin-down luminosity, $F_\gamma$ the observed gamma-ray flux
($= 3.4 \times 10^{-10} \rm erg \ cm^{-2} s^{-1}$), 
$r_s$ the radius of the bow shock, and $\theta_s = r_s/D$.
This equation suggests a reasonable density of ambient matter as
\begin{equation}
n = 1.8 \left( \frac{f_\gamma}{0.1} \right)^{-1}
 \left( \frac{\theta_s}{1''} \right)^{-2}
 \left( \frac{\upsilon}{\rm 400 km/s} \right)^{-2} \ \rm cm^{-3} \ ,
\end{equation}
although it might be higher than expected for the mid Galactic latitude
region.  We must await further multi-band optical observation to clarify
whether the extended object is a galaxy or a synchrotron nebula.

In the following we assume that C+D is not related to {\RX}/{\EG}.  The flux
of possible point sources A and B sets an upper limit of $B \gtrsim 28$ for
point sources, which is much stronger than previous limit of $V > 25.2$. This
translates to the lower limit of $f_X/f_B \gtrsim 2700$ on {\RX}. If we assume
the Reyleigh-Jeans spectrum ($B-V=-0.32$) as expected for thermal emission
from an isolated neutron star, the limit on the X-ray/V ratio becomes $f_X/f_V
>$ 5300. (However, it should also be noted that the optical spectrum of
isolated neutron stars is often not entirely the black-body, as is the 
case for the Geminga.)
This limit on X-ray/optical flux ratio further strengthens the
argument for the isolated neutron-star identification for {\RX}. 
Star counts
in such a faint magnitude are not well constrained, but a typical theoretical
model predicts about 10000--15000 stars per square degree down to $B\sim 28$
to the direction of {\RX}, with an extreme assumption that ancient white dwarfs
contribute 100\% of the dark halo of our Galaxy (Robin, Reyl\'e, \& Cr\'eze
2000; A. Robin, private communication). Then the expected star count in the
$3''$-radius error circle is about 0.03 stars, and hence a random coincidence
of a star in the error circle of {\RX} seems to be unlikely. On the other
hand, we cannot reject a possibility that the point-like objects A and/or B are
small galaxies which could not be resolved.  Keeping these caveats in mind,
we assume that one of the two is a real isolated pulsar and the physical
counterpart of {\RX}, and then discuss implications in this case.

To examine the physical connection between {\RX} and {\EG},
we should also estimate the chance probability of finding
an isolated neutron star within the error circle of {\EG}
($12'$ radius). Based on the estimations of pulsar birth rate
(e.g., Narayan \& Ostriker 1990) and supernova rate (e.g., Totani, Sato,
\& Yoshii 1996),
about $10^{8-9}$ neutron stars are typically expected to exist
in our Galaxy. Thermal emission can be observed by either 
cooling neutron star surfaces or those heated by accretion of
interstellar matters. Neuh\"auser \& Tr\"umper (1999) compiled
observational constraints on the X-ray counts of
such neutron stars detectable by the ROSAT all sky survey,
and compared them with theoretical expectations. Since neutron
stars rapidly cool down after $\sim 10^6$yr according to
the standard cooling theory of neutron stars (Friedman \&
Pandharipande 1981), the expected count of cooling neutron stars
is much smaller than that of the accreting neutron stars. The former
is $\sim$ 1--10 sr$^{-1}$ for ROSAT X-ray flux of $ \gtrsim 10^{-13}
\ \rm erg \ cm^{-2} s^{-1}$, and the expected number in the
12$'$-radius error circle of {\EG} is found to be sufficiently
small as $\sim (4$--$40) \times 10^{-5}$. The estimate for accreting neutron
stars is rather uncertain and it could be much higher than that
for cooling neutron stars, but the observational upper bound by
ROSAT is $\lesssim$ 100--1000 sr$^{-1}$ again at $f_X \gtrsim 10^{-13}
\ \rm erg \ cm^{-2} s^{-1}$, and the expected count in the
EGRET error circle should be less than $\sim$ 0.04. It should also
be noted that {\EG} is at intermediate
Galactic latitude. Then it seems
unlikely that it is heated by accretion, since the ambient gas density should
not be high and the neutron star should have high-velocity proper motion.
(The Bondi-Hoyle accretion rate scales as $\propto \upsilon^{-3}$.)

\begin{figure*}
   \begin{center}
      \FigureFile(80mm,50mm){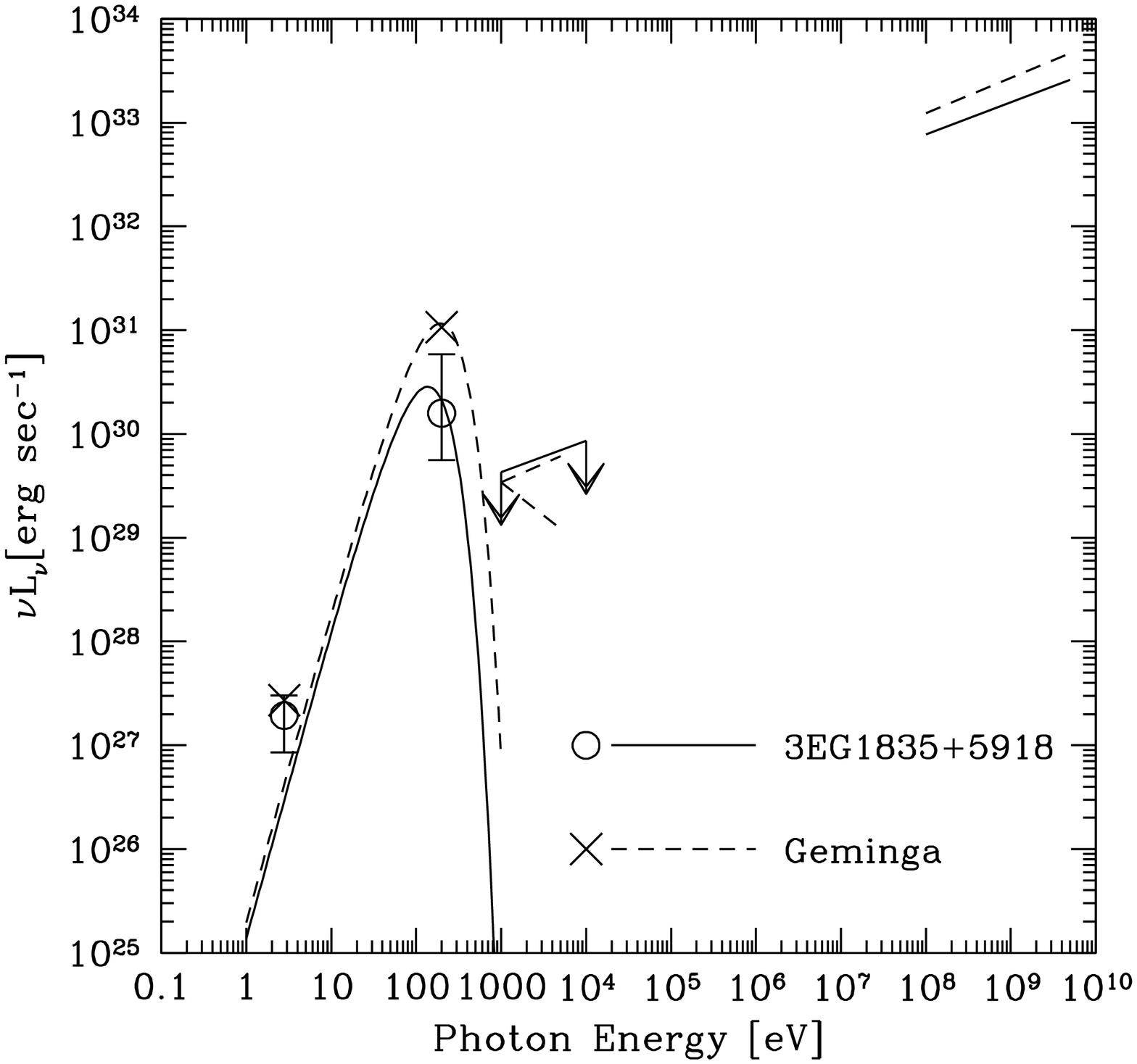}
   \end{center}
\caption{Spectral energy distribution of {\RX}/{\EG} (solid lines and open
circles) and Geminga (dashed lines and crosses), in optical, X-ray and
gamma-ray bands.  Distances are assumed to be 300pc for {\RX}/{\EG} while the
known distance of 160pc is used for Geminga. The black-body temperature is 4.1
and $5.6 \times 10^5$K for the former and the latter, respectively. The
optical flux of {\RX}/{\EG} is $B=28.0$, assuming that the source A or B is
the real counterpart of this object.  In the hard X-ray band ($>$1keV), the
upper limit from ASCA is shown for {\RX} (solid line), while the two
fits to the ASCA SIS or GIS by Halpern \& Wang (1997) are shown for Geminga
(dashed line).  
}
\label{fig:SED}
\end{figure*}

Now we discuss the broad-band spectral energy distribution of this source,
assuming that {\EG} and {\RX} are physically associated and it is a
gamma-ray pulsar.
Assuming the thermal flux of $B \sim 28$ from the neutron star surface,
we can derive thermal temperature and distance, by scaling from the
proto-type gamma-ray pulsar Geminga.  The thermal X-ray spectrum of Geminga
can be fitted with $T = 5.6 \times 10^5$K (Halpern \& Wang 1997), and its
optical $B$-band flux ($B = 26.3$, Mignani, Caraveo, \& Bignami 1998) is about
a factor of 5.8 higher than the extrapolation from the X-ray flux. Such a
trend, i.e., optical flux that is a factor of several higher than that  
extrapolated from X-ray temperature,
has been known also for another isolated neutron star, RX J185635$-$3754
(Walter \& Matthews 1997).  Assuming this factor also for {\RX}, we obtain the
black-body temperature of $T = 4.1 \times 10^5$K from the X-ray/optical flux
ratio, assuming $N_H = 1\times 10^{20} \ \rm cm^{-2}$.  If the effective
emission area of thermal radiation is the same, the optical flux in the
Rayleigh-Jeans region scales as $f \propto T / D^2$.
Therefore, the distance to {\RX} can be inferred as $D \sim 300$pc,
from the known distance of Geminga ($\sim 160$pc, Caraveo et al. 1996)
\footnote{From the X-ray temperature and bolometric thermal flux of the
Geminga pulsar (Halpern \& Wang 1997), the distance of 160pc corresponds to
the blackbody radius of $R_\infty \equiv R / [(1-(2GM/Rc^2)]^{1/2}$ =
4.6km.}.  The inferred black-body spectrum of {\RX}/{\EG} is shown in terms
of absolute $\nu L_\nu$ luminosity by the solid line in Fig. \ref{fig:SED},
and the observed gamma-ray, X-ray and optical luminosities or upper limits
are also shown by open circles or solid lines.  The error bar of the soft
X-ray data point (0.2keV) is showing the flux change when different column
densities of $N_H = 0$ or $3 \times 10^{20} \ \rm cm^{-2}$ are assumed, with
the same temperature. The upper bound on hard X-ray flux ($>$1keV) is from
ASCA observation (Mirabal et al. 2000). For comparison, the black-body
spectrum and observed luminosities of Geminga in optical, soft and hard
X-ray, and gamma-ray bands are shown by dashed lines or crosses.  A striking
similarity of the spectral energy distribution between Geminga and
{\RX}/{\EG} can be seen.  The no detection of nonthermal X-rays from {\RX} is
marginally consistent with the ASCA upper bound, if the nonthermal X-ray
luminosity is similar to that of Geminga. Assuming the distance of 300pc and
an age of the pulsar to be $\sim 3 \times 10^5$yr, which is similar to that
of Geminga, the Galactic latitude of $b = 25^\circ$ implies a reasonable
transverse proper motion of $\sim 300 (t_{\rm age} / \rm 3 \times 10^5
yr)^{-1}$ km/s.

To summarize, we found an extended source having a peculiar shape (C+D) at $B
\sim 26$ within the error circle of {\RX}, which has been suspected as an
isolated pulsar associated with {\EG}.  This source might be a pulsar nebular
or knot whose luminosity is about 5--6 orders of magnitude lower than the
gamma-ray luminosity.  However, it is statistically well possible that a
galaxy appeared by chance as the source C+D within the examined error
circle. There are two fainter possible point sources (A, B) at $B \sim 28$,
although they are not firm detections because of low signal-to-noise. It is
interesting that if either of A or B is a real counter part of
{\RX} and thermal emission from a neutron star, its optical, X, and gamma-ray
luminosities  are very similar to those of the proto-type gamma-ray
pulsar Geminga. The pulsar identification of Geminga
came from the detection of pulsation in X-ray flux, and hence it can be
imagined that the ultimate identification of the enigmatic source {\EG} will
also be brought by discovery of X-ray pulsation. On the other hand, further
deep optical study is also very important to verify whether the extended
source C+D is a galaxy or not, to confirm the possible point sources A
and/or B with higher signal-to-noise ratio, and to measure proper motion and
parallax for distance determination.

We would like to thank L. Eyer and A. Robin for useful information for star
counts. TT has been financially supported in part by the Postdoctoral
Fellowship of the Japan Society for the Promotion of Science (JSPS) for
Research Abroad.  WK is supported in part by JSPS Research Fellowship.

\begin{table}
\footnotesize
\begin{center}
\caption{Result of Observation around {\RX}}
\begin{tabular}{ccccc}
\hline \hline
object & profile & $\alpha$(J2000) & $\delta$(J2000) 
& $B$ mag  \\ 
\hline
A & point &    18h36m13.49s & 59d25m30.1s  & 29.1$^{+1.9}_{-0.7}$ \\ 
B & point &    18h36m13.90s & 59d25m29.1s  & 28.3$^{+0.6}_{-0.4}$ \\ 
C & extended & 18h36m13.87s & 59d25m31.6s  &           \\ 
D & extended & 18h36m13.93s & 59d25m32.0s  &           \\ 
C+D &        &              &              & 26.1$\pm 0.1$ \\ 
\hline \hline
\end{tabular}
\label{table:photometry}
\end{center}
\end{table}

\newpage 

\begin{figure*}
   \begin{center}
      \FigureFile(80mm,50mm){cleanimageB10x10bw.eps}
   \end{center}
\caption{The same as Fig. \ref{fig:B-30x30}, but a close-up of
the $10'' \times 10''$ region centered on {\RX}.}
\label{fig:B-10x10}
\end{figure*}

\begin{thebibliography}{200}
\bibitem[dum (1999)]{1}
Becker, R.H. \& Tr\"umper, J. 1997, A\&A, 326, 682

\bibitem[dum (2000)]{2}
Brazier, K.T.S. \& Johnston, S. 1999, MNRAS, 305, 671 (1999)

\bibitem[dum (2000)]{3}
Caraveo, P.A., et al. 1996, ApJ, 461, L91

\bibitem[dum (2000)]{4}
Schlegel, D.J., Finkbeiner, D.P., \& Davis, M. 1998, 500, 525

\bibitem[dum (2000)]{5}
Friedman, B.L. \& Pandharipande, V.R. 1981, Nuclear Physics A 361, 502

\bibitem[dum (2000)]{6}
Gehrels, N., Macomb, D.J., Bertsch, D.L., Thompson, D.J., \& Hartman, R.C.
2000, Nature, 404, 363


\bibitem[dum (2000)]{8}
Halpern, J.P. \& Wang, F.Y.-H. 1997, ApJ, 477, 905

\bibitem[dum (2000)]{9}
Hartman, R.C. et al. 1999, ApJS, 123, 79

\bibitem[dum (2000)]{10}
Harding, A.K. \& Zhang, B. 2001, ApJ, 548, L37

\bibitem[dum (2000)]{11}
Hester, J.J. et al. 1995, ApJ, 448, 240

\bibitem[dum (2000)]{12}
Hughes, J.P. et al. 2001, ApJ, 559, L153

\bibitem[dum (2000)]{13}
Kawasaki, W. \& Totani, T. 2002, ApJ in press (astro-ph/0108309)

\bibitem[dum (2000)]{14}
Mignani, R.P., Caraveo, P.A., \& Bignami, G.F. 1998, A\&A, 332, L37

\bibitem[dum (2000)]{15}
Mirabal, N., Halpern, J.P., Eracleous, M., \& R.H. Becker 2000, ApJ,
  541, 180

\bibitem[dum (2000)]{16}
Mirabal, N. \& Halpern, J.P. 2001, ApJ, 547, L137

\bibitem[dum (2000)]{17}
Narayan, R. \& Ostriker, J.P. 1990, ApJ, 352, 222

\bibitem[dum (2000)]{18}
Neuh\"auser, R. \& Tr\"umper, J.E. 1999, A\&A, 343, 151

\bibitem[dum (2000)]{19}
Reimer, O., Brazier, K.T.S., Carrami\~{n}ana, A., 
Kanbach, G., Nolan, P.L., \& Thompson, D.J. 2001, MNRAS, 324, 772

\bibitem[dum (2000)]{20}
Robin, A.C., Reyl\'e, C., \& Cr\'eze, M. 2000, A\&A, 359, 103

\bibitem[dum (2000)]{21}
Romero, G.E., in the proceedings of the Workshop on The Nature of 
Unidentified Galactic High-Energy Gamma-Ray Sources, 
Tonantzintla, October 2000. To be published by Kluwer (astro-ph/0012243)

\bibitem[dum (2000)]{22}
Seward, F.D. \& Wang, Z.-R. 1988, ApJ, 332, 199

\bibitem[dum (2000)]{23}
Slane, P. et al. 2000, ApJ, 533, L29

\bibitem[dum (2000)]{24}
Stocke, J.T., Morris, S.L., Gioia, I.M., Maccacaro, T., Schild, R.,
Wolter, A., Fleming, T.A., \& Patrick Henry, J. 1991, ApJ, 76, 813

\bibitem[dum (2000)]{25}
Totani, T., Sato, K., \& Yoshii, Y. 1996, ApJ, 460, 303

\bibitem[dum (2000)]{26}
Totani, T. \& Kitayama, T 2000, ApJ, 545, 572

\bibitem[dum (2000)]{27}
Totani, T. \& Yoshii, Y. 2000, ApJ, 540, 81

\bibitem[dum (2000)]{28}
Walter, F.M. \& Matthews, L.D. 1997, Nature, 389, 358

\bibitem[dum (2000)]{29}
Waxman, E. \& Loeb, A. 2000, ApJ, 545, L11

\end{thebibliography}
\end{document}